\documentclass[showpacs,amsmath,amssymb,twocolumn]{revtex4}

\usepackage{graphics}
\usepackage[dvips]{graphicx}

\newcommand{\gti}{\gamma i}
\newcommand{\subtext}[1]{{\mbox{\scriptsize #1}}}
\newcommand{\lo}{\hbar\omega_\subtext{LO}}
\newcounter{myeqnum}
\newenvironment{myeqnarray}{\setcounter{myeqnum}
 {\value{equation}}
 \stepcounter{myeqnum} \setcounter{equation}{0}
 
 \begin{eqnarray}}{\end{eqnarray} 
 \setcounter{equation}{\value{myeqnum}}
 
 \par \vspace{-1\baselineskip} \noindent \hspace{-0.35em}}
\newenvironment{myothereqnarray}[2]{\setcounter{myeqnum}{#1}
 \setcounter{equation}{#2}
 
 \begin{eqnarray}}{\end{eqnarray} 
 \setcounter{equation}{\value{myeqnum}}
 
 \par \vspace{-1\baselineskip} \noindent \hspace{-0.35em}}
\newcommand{\D}{\displaystyle}

\begin{document}

\title{Thermalization process of a photo-generated plasma in 
semiconductors}

\author{M.A.\ Rodr\'\i guez-Meza$^{1,2}$\footnote{Electronic address: 
mar@nuclear.inin.mx} and
J.L.\ Carrillo$^1$ }

\address{$^1$Instituto de F\'\i sica, 
  Universidad Aut\'onoma de Puebla,
  A.P.\ J--48 Puebla 72570, Puebla, M\'exico. \\ 
$^2$Instituto Nacional de Investigaciones Nucleares, 
  A.P.\ 18--1027 M\'exico D.F. 11801, M\'exico.}

\date{\today}

\begin{abstract}
The kinetics of ultra-fast processes which leads to the 
thermalization condition  of a photo-excited plasma in 
semiconductor systems is studied theoretically.  We analyze 
the time evolution of a carrier population  generated by a 
finite optical pulse, from the beginning of the pulse until 
the time in which the carrier population reaches a 
quasi-equilibrium condition.  We calculate the  energy 
fluxes caused by the main interaction mechanisms along the 
different stages  the system passes through. Our analysis is 
done by using a set of non-linear rate equations which govern 
the time evolution of the carrier population in the energy 
space. We consider the main interaction mechanisms, including 
dynamic screening and phonon population effects. 

\vspace{0.25\baselineskip}

\vspace{0.75\baselineskip}

Se estudia la
cin\'etica de los procesos ultra r\'apidos que llevan
a la condici\'on de termalizaci\'on de un plasma fotoexcitado
en sistemas semiconductores.
Anali\-zamos la evoluci\'on temporal de una poblaci\'on 
generada por un pulso \'optico finito, desde el comienzo
del pulso hasta el tiempo en el que la poblaci\'on alcanza
una condici\'on de cuasi equilibrio.
Calculamos los flujos de energ\'\i a causados por los mecanismos
principales de interacci\'on a lo largo de las diferentes etapas
por las que pasa el sistema.
Hacemos nuestro an\'alisis usando un conjunto no lineal de 
ecuaciones de raz\'on de cambio que gobiernan la evoluci\'on
temporal de la poblaci\'on de portadores en el espacio de 
energ\'\i as.
Consideramos los mecanismos principales de interacci\'on,
incluyendo el apantallamiento din\'amico y efectos
de poblaci\'on de fonones.

\vspace{0.25\baselineskip}
\noindent {\em Descriptores}: plasma fotoexcitado; termalizaci\'on;
procesos ultra r\'apidos en semiconductores

\end{abstract}


\maketitle

\section{Introduction}
The relaxation processes of 
photo-generated plasma systems 
in semiconductors exhibit two characteristic stages. The 
first of them, commonly called the thermalization process, is 
mainly governed by the rapid interactions, namely, the 
electron-electron ($e$-$e$) and the electron-optical phonon 
interactions. In this stage the carrier distribution function 
(CDF) is far from equilibrium and theoretical approximations 
for this, based on 
small
displacements from the equilibrium 
are poor approaches to describe the kinetics of these 
processes. This stage ends when the interaction mechanisms 
randomize the energy and momentum in the carrier population. 
This allows the CDF to reach a condition in which it is 
possible to define an effective temperature for the carrier 
system, i.e., the CDF acquires a shape similar to that of an 
equilibrium one. The second stage of the relaxation is the so 
called, cooling 
process, and 
is mainly ruled by the slow 
interactions in the system, namely, electron-phonon ($e$-ph) 
scattering and recombination. This process has been 
extensively studied since the pioneering works of the late 
60's and 70's decades\cite{shah,hess}. In this stage, the time 
evolution of the quasi-equilibrium CDF can be described by 
means of simplified evolution equations for time dependent 
effective temperature and chemical potential\cite{alfanobook}. 

There have been in the literature reported 
theoretical and experimental studies on the thermalization 
process, however it is still not well 
understood\cite{hess,heiner,marjl1,marjl2,sun}. This is 
because the
thermalization 
process is a stage dominated by
transient effects in a 
far from equilibrium  system. The understanding of the 
transient processes occurring in photo-generated carrier 
populations is of great relevance because it would 
allow a deeper physical insight on the dynamical effects of interaction 
mechanisms upon observable properties of the system. In addition, it 
could provide information, as well as a theoretical framework, 
to investigate some particular ultra-fast phenomena like the 
kinetics of thermo-transport and the kinetics of the coherent 
control of quantum states in mesoscopic semiconductor 
systems\cite{cartagenavilnius,potz,wehner}.

Thermalization depends mainly on the efficiency of the $e$-$e$ 
scattering to redistribute the excess energy given to the 
system by the external sources. Once thermalized, the 
electronic system relaxes by dissipating the energy in excess 
into the lattice and by emitting radiation. Associated to each 
one of the stages of the relaxation process, there is a 
characteristic time.  The first one, the thermalization time, 
is an effective time determined by the intrinsic characteristic times of 
the rapid interaction mechanisms within the system and the second 
one is a characteristic time determined by the interaction 
mechanisms of the system with the surroundings. Thermalization 
and cooling processes also depend on other features of the 
excitation, for instance the excitation time, i.e., the time 
along which the carrier generation is produced or the 
time the
perturbation 
remains switched on. The thermalization and cooling times 
notoriously change if the external perturbation remains 
switched on over times longer than the $e$-$e$ collision time, 
or if the perturbation switching on occurs adiabatically.

We present in this paper a detailed study on the ultra-fast 
processes which lead to the thermalization condition in a 
photo-excited electron gas in polar semiconductors. We assume 
the semiconductor is excited by a pulsed laser. We define two 
quantities on which our discussion is based, namely, the 
thermalization time and the thermalization 
temperature\cite{marjl1}. We analyze how these quantities 
depend on some external variables such as the time duration 
of the laser pulse, the energy of excitation, the carrier 
concentration, and the lattice temperature.

\section{Theory}
To study the thermalization process of a photo-generated 
electron gas in a bulk semiconductor we use a set of rate 
equations we have developed to investigate several physical 
situations\cite{marjl1,marjl2}. These equations describe the 
evolution of the electron population in semiconductors under 
several general conditions, and consider the most important 
interaction mechanisms, including screening and phonon 
population effects. Our theoretical scheme is based in the 
following formalism.

Let us start by defining the carrier population in a 
volume element in the $({\bf r},{\bf k})$-space
or $\mu$-space, composed of the carrier position ${\bf r}$ and
the carrier wavevector ${\bf k}$,
\begin{myeqnarray}
  \eta({\bf r},{\bf k},t) \,=\, 
  f({\bf r},{\bf k},t)\,\frac{V}{4\pi^3}\,d{\bf r}\, d{\bf k}
\end{myeqnarray}
where $f({\bf r},{\bf k},t)$ is the carrier distribution 
function in the $\mu$-space and $V$ is the crystal volume.

In a similar way, for the optical phonon population we can 
write
\begin{myothereqnarray}{\value{equation}}{1}
  N_j({\bf q},t) \,=\, 
  g_j({\bf q},t)\,\frac{V}{(2\pi)^3}\,d{\bf q} \quad ,
\end{myothereqnarray}
where $j$ labels the branch, ${\bf q}$ is the phonon wave 
vector and $g_j({\bf q},t)$ is the phonon distribution 
function. In the following we will use the index $\alpha$ to 
denote the couple $(j,{\bf q})$. So that $N_\alpha$ will
denote the phonon population in the mode $\alpha$. 

An electron in a semiconductor is characterized by its position
${\bf r}$, its wavevector ${\bf k}$, and the index of the energy band.
An electric field introduces a preferred direction.
By assuming an homogeneous, isotropic system, and
for null applied electric 
field, the only relevant variable is the energy, as is the
case of cubic semiconductors like GaAs\cite{marjl2,shah1999}. Therefore,
we can write 
for the carrier and phonon populations
\begin{myeqnarray}\label{p2000_eq_02}
  \eta(\epsilon,t)&=&f(\epsilon,t)\,\frac{V\,dS\,d\epsilon}
  {4\pi^3 \vert \nabla_{\bf k}\epsilon({\bf k}) \vert} 
	\quad,\\[.1in]
  N_\alpha(t)&=&g_j({\bf q},t)\,\frac{V}{(2\pi)^3}\,d{\bf q} 
	\quad ,
\end{myeqnarray}
where $dS$ is a surface element on the surface of constant 
energy $\epsilon$.

Now we wish to establish the equations which govern the time 
evolution of these quantities. In order to do that we assume 
the following. 
Transport and optical properties of cubic semiconductors like GaAs
are explained in terms of their band structure. The band structure
of a cubic-model semiconductor is composed of one conduction band
with three sets of minima, and three valence bands. The minima of the 
conduction band are located at the $\Gamma$ point ($k=0$), at the
$L$ points [${\bf k}=(\pi/a_0,\pi/a_0,\pi/a_0)$, $a_0$ being the lattice
constant], and along the $\Delta$ lines ${\bf k}=(k,0,0)$. The tops of
the valence bands are located at $\Gamma$ point. Two of them are degenerate
at this point and the other is separated by spin-spin interaction.
Then,  
we suppose a band structure composed of the 
valley $\Gamma$ separated from the valley $L$ by an energy 
$\Delta$. This model is appropriate for the description of 
photoexcited semiconductors where the transitions involve only 
the center of the Brillouin zone\cite{shah1999}.
We assume also that the continuum of states of the 
valley is partitioned into a set of discrete intervals of 
energy $\Delta\epsilon$. For simplicity, and in order to have
a direct reference we choose $\Delta\epsilon$ 
to be the longitudinal optical (LO) phonon of energy $\lo$. 
This choice  is not an essential assumption for the 
development of our procedure\cite{marjl2}.
 
Instead of 
establishing
evolution equations for the quantities
$\eta(\epsilon,t)$, we shall set up the evolution equations 
for these quantities integrated in the interval of range
$\Delta\epsilon=\lo$. So, we have
\begin{equation}\label{p2000_eq_03}
\eta_{\gti}(t) \,=\, f_\gamma(\epsilon_i,t)\,d_{\gti} \quad ,
\end{equation}
where $\gamma=\Gamma$,$L$ labels the valleys and $i$ the 
levels of energy $\epsilon_i=i\Delta\epsilon$; 
\makebox{$i=0,1,2,\ldots$}\,, and
\begin{eqnarray}
  d_{\gti} \,\,=\,\, \int_{\Delta\epsilon}\,\,
      D_\gamma(\epsilon_i)\,d\epsilon \,\,=\,\, 
      \int_{\Delta\epsilon}\,\,\frac{ V\, dS\, d\epsilon }
      { 4\pi^3 \vert \nabla_{\bf k}\epsilon({\bf k}) \vert }
\end{eqnarray}
is the number of energy states in the interval characterized 
by set of indexes $(\gti)$. Obviously it depends on the 
density of states $D_\gamma(\epsilon)$ in the respective 
valley. 

We obtain the evolution equation for the carrier population 
in the different levels in the energy space by using the 
conservation of the electron number, thus we can write
\begin{equation}\label{p2000_eq_04}
  \frac{ d\eta_{\gti} }{ dt } =
    \sum_m ( b_m - a_m ) + G_{\gti} - R_{\gti}   \quad ,
\end{equation}
where $a_m$ ($b_m$) is the flux out from (into) the level 
$\gti$ due to the interaction mechanism labeled by  $m$. The 
photo-generation of carriers is accounted for $G_{\gti}$ while 
$R_{\gti}$ denotes the recombination rate. The rate equations
(\ref{p2000_eq_04}) govern the evolution of the carrier system
and form a set of non linear coupled differential equations where
the fluxes depend on the carrier populations
\begin{eqnarray*}
 a_m &=& a_m(\eta_{\gamma i-1}, \eta_{\gamma i}, \eta_{\gamma i+1}) \\
 b_m &=& b_m(\eta_{\gamma i-1}, \eta_{\gamma i}, \eta_{\gamma i+1})
\end{eqnarray*}

\subsection{Scattering mechanisms}
We now particularize our treatment to the case of polar 
semiconductors. However, the adaptation of the theory 
necessary to deal with covalent semiconductors is almost 
direct. The evaluation of the probabilities associated to the 
collision mechanisms can be done by using the straightforward 
first-order perturbation theory. The expressions obtained by 
the use of the Fermi Golden rule for the transition 
probabilities due to the different interaction mechanisms, can 
be found somewhere else. The derivation of some of them and 
modifications of these expressions according with our 
theoretical framework, is almost direct, here we just discuss 
those details which in our opinion might need some
clarification. Full details can be found in 
Ref.\ \cite{marjl2}.

\subsubsection{Electron-longitudinal-optical phonon 
interaction (polarization potential)}
When a carrier undergoes an electron-logitudinal-optical phonon
interaction makes a transition to the neigbour levels ($i\pm 1$) in 
the same band. Transition to different bands (inter valley) 
in which an optical phonon is
participating is due to a different potential, the optical deformation 
potential\cite{shah1999}. For photoexcited semiconductors we are interested in 
the transitions occur only at the center of the Brillouin zone, and
these inter valley transitions are neglected\cite{shah1999}. The expression for
the fluxes due to the electron-logitudinal-optical phonon 
interaction is derived in \cite{marjl2} and are based on
the matrix elements derived elsewhere, see for example \cite{Conw}.

The fluxes can be written as\cite{marjl2,Conw,Yoff}
\begin{myeqnarray}
  a &=& \eta_{\gti} \nu_\subtext{op}^{\gamma\pm}(\epsilon_i)
        \left (
        1 - \frac{ \eta_{\gamma i\mp1} } { d_{\gamma i\mp1} }
        \right ) \\[.05in]
  b &=& \left ( 1 - \frac{ \eta_{\gti} }{ d_{\gti} } \right )
            \nu_\subtext{op}^{\gamma\pm}( \epsilon_i)\,\,
            \eta_{\gamma i\pm1}  \quad ,
\end{myeqnarray}
where
\begin{eqnarray}
  \nu_\subtext{op}^{\gamma\pm}(\epsilon) &=&
    \frac{ \sqrt{ m_\gamma } e^2 \lo }{ \sqrt{2} \hbar^2 }
    \left (
    	\frac{1}{ {\cal E}_\infty } - 
		\frac{1}{ {\cal E}_\subtext{s} }
    \right )
    \left ( N_q + \frac{1}{2} \pm \frac{1}{2} \right )
	\times \nonumber \\[.1in]
  & & 
    \frac{1}{ \sqrt{ \epsilon } }  
	S_\subtext{LO}^\gamma 
	\ln{ \left [
            	\frac{ 1 + \sqrt{ 1 \mp 
			\frac{ \D \lo }{ \D \epsilon } } }
                   { \pm 1 \mp \sqrt{ 1 \mp 
			\frac{ \D \lo }{ \D \epsilon } } }
            \right ] } \quad .
\end{eqnarray}
${\cal E}_\infty$ and ${\cal E}_\subtext{s}$ are the static 
and optical dielectric constants, respectively. $m_\gamma$ 
is the effective mass in the $\gamma$ valley, $e$ is the 
electron charge and $\hbar$ is the Planck constant. The upper 
(lower) sign is for emission (absorption). The screening 
effects in the $e$-ph interaction are included in the factor
$S_\subtext{LO}^\gamma$, which in the random-phase 
approximation is given by\cite{Yoff}
\begin{displaymath}
  S_\subtext{LO}^\gamma \,=\,\, \frac{ N_\gamma }{ N }
      \left (
         1 + \left ( \frac{ N_\gamma }{ N_\gamma^\subtext{c} } 
		\right )^2
      \right )^{-1}
\end{displaymath}
where $N$ and $N_\gamma$ are the total and $\gamma$ valley 
carrier concentration respectively, $N_\gamma^\subtext{c}$ is 
the threshold value for the concentration in the $\gamma$ 
valley at which the screening becomes important. This 
critical value for the carrier concentration is given by
\begin{displaymath}
 N_\gamma^\subtext{c} \,=\,\,
 \frac{ {\cal E}_\infty m_\gamma (\lo)^3 }
 { 3^{3/2} 8 \pi e^2 \hbar^2 k_\subtext{B} T_\subtext{e} }
      \quad ,
\end{displaymath}
where $T_\subtext{e}$ is the electronic temperature and 
$k_\subtext{B}$ is the Boltzmann's constant.

\subsubsection{Electron-electron interaction}
The electron-electron interaction gives the nonlinear character of the 
Eqs.\ (5) and is one of the most difficult interactions 
to take into account. We adopt the Debye-H\"uckel screened potential
to describe the $e$-$e$ interaction. We see that the scattering processes in
which the magnitude of exchanged momentum is small are the most likely,
because de probability diminishes as $q^{-4}$. Also, on the average, an 
electron in the valley $\Gamma$ exchange approximately $\hbar\omega_{LO}$ of
energy\cite{marjl2}. Therefore, in our energy levels scheme, 
carriers make transitions to the neighbour
levels $i\pm 1$ in the same band due to the $e$-$e$ scattering. 
For more details see Refs.\ \cite{marjl2,marphasetr}.

The fluxes are given by\cite{marjl2}
\begin{myeqnarray}
  a &=& \eta_{\gti} Z^\gamma
    \left [
      \left ( 1 - 
	\frac{ \eta_{\gamma i-1} }{ d_{\gamma i-1} } \right )
      \sum_{\gamma' i'}\,\, \eta_{\gamma' i'}
      \left ( 1 - 
	\frac{\eta_{\gamma' i'+1}}{d_{\gamma'i'+1}} \right )
      \nonumber \right. \\[.05in]
    & & \left. \mbox{} +
      \left ( 1 - 
	\frac{ \eta_{\gamma i+1} }{ d_{\gamma i+1} } \right )
      \sum_{\gamma' i'}\,\, \eta_{\gamma' i'}
      \left ( 1 - 
	\frac{\eta_{\gamma' i'-1}}{d_{\gamma'i'-1}} \right )
    \right ]  \\[.2in]
  b &=& \left ( 1 - 
	\frac{ \eta_{\gti} }{ d_{\gamma i} } \right ) Z^\gamma
    \left [
      \eta_{\gamma i+1}
      \sum_{\gamma' i'}\,\, \eta_{\gamma' i'}
      \left ( 1 - 
	\frac{\eta_{\gamma' i'+1}}{d_{\gamma'i'+1}} \right )
      \nonumber \right. \\[.05in]
    & & \left. \mbox{} +
      \eta_{\gamma i-1}
      \sum_{\gamma' i'}\,\, \eta_{\gamma' i'}
      \left ( 1 - 
	\frac{\eta_{\gamma' i'-1}}{d_{\gamma'i'-1}} \right )
    \right ]
\end{myeqnarray}
where Pauli exclusion principle has been taken into account and
\begin{equation}\label{eq_16}
  Z^\gamma \,\,=\,\,
    \frac{ e^2 
	\sqrt{ \pi m_\gamma k_\subtext{B} T_\subtext{e} } }
         { 2^2 \hbar^2 {\cal E}_\infty N }
    \left [
      \frac{ 1 }{ 1 + 
	\frac{ \D N }{ \D N_\subtext{ee}^\gamma } }
    \right ]  \quad ,
\end{equation}
The square brackets factor takes into account the screening 
effects. These become important when the carrier 
concentration $N$ reaches a
critical value $N_\subtext{ee}^\gamma$ given by
\begin{displaymath}
  N_\subtext{ee}^\gamma \,\,=\,\,
    \frac{ 4 m_\gamma {\cal E}_\infty 
	( k_\subtext{B} T_\subtext{e} )^2 }
         { \pi^2 \hbar^2 e^2 }   \quad .
\end{displaymath}

The expression for the total probability, Eq.\ (\ref{eq_16}), 
is an heuristic useful expression, that allows us to determine 
in an easy way the ranges of carrier concentration and 
electronic temperature, in which the energy exchange through 
$e$-$e$ scattering is the dominant mechanism in the kinetics 
of the system.

\subsection{Generation and recombination}
The dynamics of carriers under intense laser irradiation has 
been described by Ferry\cite{Fer2}. Here we adopt the 
following procedure to describe the generation and 
recombination processes. The rate equation which governs the
effect of the generation and recombination processes on the 
CDF can be written as
\begin{equation}
  \frac{ d\eta_{\Gamma i} }{ dt } \,\,=\,\,
       G_{\Gamma i} - R_{\Gamma i}
\end{equation}
the first r.h.s term represents the photo-generation of 
carriers and the second one represents the recombination 
processes. In direct gap semiconductors the photo excitation
involve the top of the valence band and the bottom of the conduction
band which are at the $\Gamma$ point. For indirect gap semiconductors
it is necesary the participation of a phonon in order to conserve
momentum. 
In this report we are interested in direct gap semiconductors,
therefore,
we shall assume only generation of carriers to the 
$\Gamma$ valley, therefore, the generation term can be written
as\cite{marjl2}
\begin{eqnarray}
  G_{\Gamma i} \,\,=\,\, G_p(t)\, \delta_{i,i_p}
     \left ( 1 - 
	\frac{ \eta_{\Gamma i} }{ d_{\Gamma i} } \right )
     \left ( 1 - \sigma 
	\frac{ \eta_{\Gamma i} }{ d_{\Gamma i} } \right )
\end{eqnarray}
where $i_p\Delta\epsilon$ is the excitation energy 
(measured from the bottom of the conduction band). $G_p(t)$ 
is proportional to the rate of generation, i.e., the number 
of excited carriers per unit time given by
\begin{displaymath}
  G_p(t) \,\,=\,\, G_p^* G(t) \,\,=\,\,
         \frac{ P_\subtext{L} }{ A d \hbar 
	\omega_\subtext{L} } G(t) \quad ,
\end{displaymath}
where $P_\subtext{L}$ is the laser power, 
$\hbar\omega_\subtext{L}$ is the photon energy, $d$ is the 
penetration length, and $A$ is the area of illumination. 
The dimensionless function $G(t)$ is conveniently chosen in 
order to model a given experimental condition. For instance, 
for the case of a steady state $G(t)=1$. The factor 
$\sigma$ is given by $\bar{f}/f$ where $\bar{f}$ is the hole distribution,
and in 
general is a function which depends upon 
$\epsilon$ and $t$. In steady state processes $\sigma$ 
accounts for the degree of compensation in the material.

The rate of recombination can be expressed as 
$R=R^\subtext{R}+R^\subtext{N}$, i.e., it is the addition of 
the radiative and non radiative components of the 
recombination.

In this work we assume $R_{\Gamma i}=w\eta_{\Gamma i}$ where
$w$ is defined as
\begin{eqnarray}
  w \,\,=\,\, \langle\frac{1}{ \tau_\subtext{rec} }\rangle 
	\,\,=\,\,
    \frac{
      \int\,\,
        \frac{\D 1}{ \D \tau_\subtext{rec} }
        f( \epsilon ) \sqrt{ \epsilon }\, d\epsilon }
    { \int\,\, f( \epsilon ) \sqrt{ \epsilon }\, d\epsilon }
	\quad ,
\end{eqnarray}
and $\tau_\subtext{rec}$ has to be calculated for the 
pertinent kind of recombination.

\subsection{Phonon population effects}
We have mentioned that a high population of phonons can 
produce some important effects and these effects are more 
notorious in the case of the LO phonons\cite{cocevar}. 
However, the number of LO modes 
is
limited by the magnitude 
of the electron wave vector. These limits can be easily 
obtained by applying the energy and momentum conservation 
conditions to a transition in which an electron with an 
energy near the maximum energy $\epsilon_\subtext{max}$ 
absorbs a LO phonon. In this way one obtains
\begin{displaymath}
  q_\subtext{min} \,\,=\,\,
     \frac{ \sqrt{ 2 m_\gamma 
	\epsilon_\subtext{max} } }{ \hbar }
     \left [
       \sqrt{  1 + \frac{ \lo }{ 
	\epsilon_\subtext{max} } } - 1
     \right ] \quad \mbox{}
\end{displaymath}
and
\begin{displaymath}
  q_\subtext{max} \,\,=\,\,
     \frac{ \sqrt{ 2 m_\gamma 
	\epsilon_\subtext{max} } }{ \hbar }
     \left [
       \sqrt{  1 + 
	\frac{ \lo }{ \epsilon_\subtext{max} } } + 1
     \right ] \quad .
\end{displaymath}
Of course in this energy scheme it is not possible to know 
the wave vector of the absorbed phonon. In order to describe 
phonon population effects we need to link the carrier system 
to the phonon population. To this end we use the following  
evolution equation for the whole phonon population 
$N_\subtext{LO}$
\begin{eqnarray}\label{p2000_eq_11}
  \frac{ dN_\subtext{LO} }{ dt } &=&
     \frac{1}{ U } \sum_{\gti}\,\, \eta_{\gti} 
     \left [
        \nu_\subtext{op}^{\gamma +} ( \epsilon_i)
          \left ( 1 - 
	\frac{ \eta_{\gamma i-1} }{ d_{\gamma i-1} } \right )
     \right. \nonumber \\[.05in]
     & & \left. \mbox{} -
        \nu_\subtext{op}^{\gamma -} ( \epsilon_i )
          \left ( 1 - 
	\frac{ \eta_{\gamma i+1} }{ d_{\gamma i+1} } \right )
     \right ] 
     \nonumber \\ & & 
     -
     \frac{ N_\subtext{LO} - 
	N_\subtext{LO}^\subtext{eq} }{ \tau_\subtext{LO} }
\end{eqnarray}
where $U$ is the number of permitted modes per unit volume,
\begin{displaymath}
  U \,\,=\,\, \int_{q_\subtext{min}}^{q_\subtext{max}}\,\,
                        \frac{ dq }{ (2\pi)^3 }   \quad .
\end{displaymath}
$N_\subtext{LO}^\subtext{eq}$ is the equilibrium total LO 
phonon population and $\tau_\subtext{LO}$ is its time life.

The set of equations (\ref{p2000_eq_04}) and 
(\ref{p2000_eq_11}) with the respective expressions for the 
fluxes are the basis of this model. These equations can be 
easily extended to represent a more general situation. For 
instance, the hypothesis of parabolic valleys can be changed 
to allow the description with a more realistic band structure. 
The solution to these rate equations can be considerably 
simplified by the use of effective collision frequencies 
(ECF). These frequencies are the average over the band of the 
scattering frequencies defined for each of the energy levels
and which are the factors appearing in the fluxes.
We would like to notice here that the information about the 
band structure and other symmetries of the system is 
contained in the ECFs. Thus, this approach can be used to 
study ultra-fast phenomena in quantum wells, superlattices 
and heterostructures, as well as bulk semiconductors 
systems\cite{marjl2,lilia96}.

\section{Normalized rate equations}
For not very high carrier concentration we can neglect the 
inter valley transitions. The threshold in which we can
neglect inter valley transitions depends on the material,
for GaAs this threshold is $1\times 10^{18}$ 
cm$^{-3}$\cite{shah1999}. If additionally one assumes a 
non-degenerate electron gas and use the ECFs as defined 
before, the rate equations (\ref{p2000_eq_04}) and 
(\ref{p2000_eq_11}) become notoriously simplified. The use of 
the ECFs in our description is in fact justified by the smooth 
behavior of the scattering probabilities as function of the 
energy. Hence, under these conditions the rate equations can 
be cast 
into
\begin{myeqnarray}\label{p2000_eq_14}
  \frac{ d\chi_i }{ dt } &=&
    \nu_\subtext{o}^+ ( \chi_{i+1} - \chi_i ) +
    \nu_\subtext{o}^- ( \chi_{i-1} - \chi_i ) 
    \nonumber \\
  & & \mbox{} +
    ZN_\subtext{max}\chi( \chi_{i+1} - 
	2\chi_i + \chi_{i-1} ) \nonumber \\
  & & \mbox{} +
    ZN_\subtext{max}\chi_0 ( \chi_i - \chi_{i-1} )
    \nonumber \\
  & & \mbox{} +
    g_p \delta_{i,i_p} - w\chi_i \quad ,
\end{myeqnarray}
for $i\neq0$ and for $i=0$ we have
\begin{myothereqnarray}{\value{equation}}{1}
  \frac{ d\chi_0 }{ dt } &=&
    \nu_\subtext{o}^+ \chi_1 - \nu_\subtext{o}^- \chi_0
    \nonumber \\
  & & \mbox{} +
    ZN_\subtext{max} \chi ( \chi_1 - \chi_0 ) +
    ZN_\subtext{max} \chi_0\chi_0  \nonumber \\
  & & \mbox{} +
    g_p \delta_{0,i_p} - w \chi_0  \quad .
\end{myothereqnarray}
For the phonon population in excess,
$N_\subtext{LO}^\subtext{exc}
 = N_\subtext{LO}-N_\subtext{LO}^\subtext{eq}$, we have
\begin{myothereqnarray}{\value{equation}}{2}
  \frac{ dN_\subtext{LO}^\subtext{exc} }{ dt } &=&
    \frac{1}{u}
    \left [ \nu_\subtext{o}^+ ( \chi - \chi_0 ) - 
	\nu_\subtext{o}^- \chi \right ] 
    - \xi N_\subtext{LO}^\subtext{exc}
\end{myothereqnarray}
where the populations have been normalized to the maximum 
reachable carrier concentration $N_\subtext{max}$, i.e.,
$\chi_i=\eta_i/N_\subtext{max}$; $\chi = \sum_i \chi_i $;
$u=U/N_\subtext{max}$; and $g_p = G_p/N_\subtext{max}$.
The ECF $\nu_\subtext{o}^\pm$ has 
just the dominant term of the $e$-LO phonon interaction, and 
$\xi=1/\tau_\subtext{LO}$. Notice the differences between the 
rate equation for $i=0$ and $i\neq0$.

\section{Results and discussion}
For the sake of brevity, from here 
on,
we will refer as 
carrier distribution function to the set of values 
$\{\chi_{i}\}$  of the carrier population at the different 
energy intervals on the conduction band. The link of this 
distribution with the actual out of equilibrium CDF is given 
by the expression (\ref{p2000_eq_03}). We start our 
discussion by defining two 
physical quantities inherent to the time evolution of the 
photo-generated carrier population.
More specifically, we wish to characterize by means of these
physical quantities, the stage in which the system reaches
the thermalization condition.
These quantities are the thermalization time 
$t^{*}$ and the thermalization temperature $T^{*}_{e}$\cite{marjl1}. 
Our 
definition of these quantities intend to be phenomenologically 
amenable. Thus, $t^{*}$ is defined as the time interval, 
measured from the begining of the laser pulse 
that the carrier system requires
to reach a distribution 
shape which can be fitted by means of a single exponential 
function,
\begin{equation}\label{p2000_eq_15}
\chi_i = A\exp{(-i\Delta\epsilon/{k_{B}T^{*}_{e}})}
\end{equation}
Here A is a normalization constant. The quantity  $T^{*}_{e}$, 
which makes the fitting possible, defines the thermalization 
temperature. 
This last definition closely resembles the way in which the 
carrier temperature of a hot electron system is experimentally
determined\cite{shah}.
We proceed by numerically integrating  
Eqs.\ (\ref{p2000_eq_14}-c) under different physical conditions 
and seeking the effects of the external variables on $t^*$ and 
$T_e^*$. 
Starting at the pulse begining, the carrier population generated
in the conduction band evolves due to the collision mechanisms
according to the rate equations (\ref{p2000_eq_14}-c).
At each step of the numerical integration we are able to calculate 
the CDF and the corresponding values of the ECFs. 
In the initial steps of the numerical integration, in general
the CDF differs clearly of the shape of a thermalized distribution,
i.e., a Boltzmann factor. Within subsequent iterations the CDF gradually
addopts a decreasing shape, which eventually, at $t=t^*$, admits
a fitting by means of a simple decreasing exponential function.
The exponent in that event is inversely proportional to 
$T_e^*$\cite{shah,hess,marjl1,marjl2}. 
This is the way in which we proceed to evaluate $t^*$ and
$T_e^*$.

In order to make concrete calculations we consider the well 
known values of the electronic band structure, phonon dispersion relations
and material parameters of GaAs. 
For example, $\hbar\omega_{LO}=36$ meV and the energy extent of
the conduction band at the $\Gamma$ point is 1 eV. This means
that the necesary number of energy levels is 28. 
Other GaAs material parameters
can be found in \cite{marjl2,shah1999} and references there in.
At $t=0$ in an empty conduction band a laser pulse of duration $t_p$
injects $g_p$ electrons per unit time 
with an energy $i_p$ in units of $\Delta \epsilon$
and above of the bottom of the band.

\begin{figure}
\includegraphics[width=2.75in]{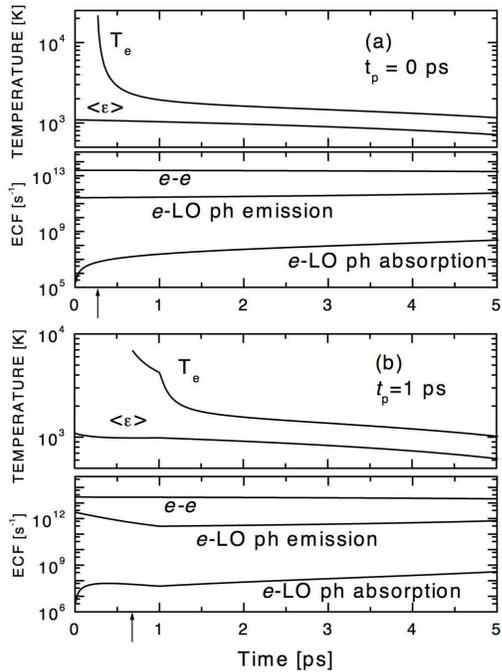}
\caption{Time evolution of the effective carrier temperature 
$T_e$ and the kinetic temperature obtained from the mean 
energy in units of K, $2\langle\epsilon\rangle/3k_B$. The 
time evolution of the main ECFs are shown in the lower part.
The pulse duration is (a) $t_p= 0$ ps and (b) $t_p=1$ 
ps.}
\end{figure}
Before starting 
our analysis of the thermalization process we discuss briefly 
the physical suitability of our carrier temperature 
definition. By means of expression (\ref{p2000_eq_15}) we have 
defined at time $t^*$ the thermalization temperature $T_e^*$.  
At subsequent times one also might use this procedure to 
calculate the carrier temperature $T_e(t)$.  We have studied 
the time evolution of this quantity under various different 
physical conditions. Our results show a good  agreement with 
experimental data\cite{marjl2}. Firstly we analyze the 
effect of the pulse duration on the time evolution of the 
carrier temperature $T_e$, as well as on the main interaction 
frequencies, i.e., $e$-$e$, $e$-ph emission, and $e$-ph 
absorption; $z$,  $\nu^+$, and $\nu^-$ respectively. In 
Fig.\ 1(a) we show $T_e(t)$ for an instantaneous pulse $t_p=0$ 
ps which excites $N=10^{17}$ cm$^{-3}$ electrons at an energy
level $i_p=4$. We have also plotted the behavior of the 
kinetic temperature defined by $2\langle\epsilon\rangle/3k_B$. 
We show the 
behavior of these quantities since the pulse starts, until 
times well above the thermalization condition is reached. We assume in 
this calculation a lattice temperature $T_L=10$ K and take 
from the literature a commonly used value of the damping 
constant for the LO-phonons $\tau_{LO}=12$ ps. Notice that the 
resulting thermalization time is $t^*=0.26$ ps (indicated by 
an arrow in the figure). In the lower 
part of Fig.\ 1(a) appears the time evolution of the main ECFs 
during this relaxation process. In Fig.\ 1(b) we show these 
quantities for a pulse of finite duration $t_p=1$ ps. In this 
case we obtain $t^*=0.65$ ps (also indicated by an arrow in 
the figure). From the comparison of these 
figures we may conclude that the pulse duration has a clear 
influence on the general characteristics of 
these
quantities, 
namely, the pulse duration modifies $t^*$ and $T_e^*$. Thus, 
this is one of the external variables which determines the 
interaction mechanisms and their relative importance in the 
thermalization process of the CDF. The evolution of the main 
ECFs  provide detailed information, at every stage of the 
relaxation process, about the relative incidence of the main 
collision mechanisms in the kinetics of the carrier system. We 
wish to recall here that we are considering in our treatment 
screening and phonon population effects. We notice that the 
$e$-$e$ ECF varies rather slightly along the period of time 
considered, but the ECFs corresponding to $e$-ph interaction 
exhibit more pronounced variations.  Notice the 
semi-logarithmic scale of the graphics. In Fig.\ 1(b) one can 
observe a clear abrupt change in the time evolution of the 
electronic temperature at the time when the pulse ends. We now 
restrict our analysis to the thermalization process and, on 
the basis of the behavior of the ECFs in this stage, we 
discuss the effect of $t_p$ and other external parameters on 
$T_e^*$ and $t^*$. 
      
\begin{figure}
\includegraphics[width=2.75in]{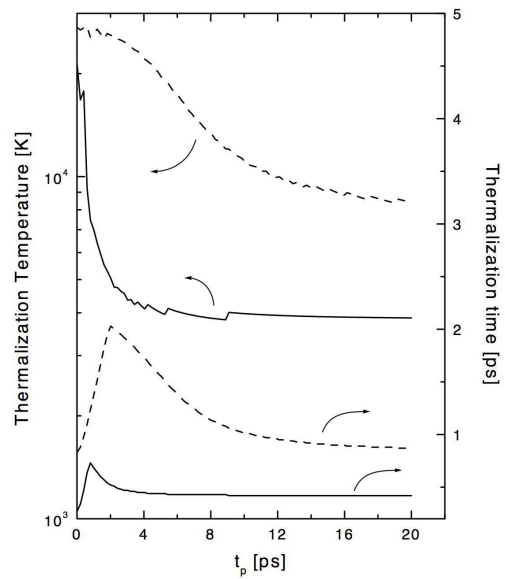}
\caption{Temperature and time of thermalization vs pulse 
duration for two values of the excitation energy level. 
Continuous lines correspond to $i_p=4$ and dashed lines to 
$i_p=10$. The carrier concentration is $N=10^{17}$ cm$^{-3}$.
The scale for $t^*$ is at the right axis of the 
graph.}
\end{figure}
Fig.\ 2 shows the changes which the pulse duration 
induces on $T_e^*$ and $t^*$. We consider here a carrier 
population $N=10^{17}$ cm$^{-3}$, a lattice temperature 
$T_L=10$ K, and two energy levels of excitation $i_p=4$ and 
$i_p=10$.  Notice that the scale for $t^*$ appears on the 
right hand side of the graphs. We observe that both $T_e^*$ 
and $t^*$ reach higher values for $i_p=10$ than they do for 
$i_p=4$, this is so all along the interval of $t_p$. This 
result is easily explained if we realize that for $i_p=10$ 
the carrier system receives a higher excess energy than 
it does for $i_p=4$, it leads the carrier system to 
reach, comparatively, a higher value of $T_e^*$ and 
due to the also relatively larger number of energy states
accesible to the carriers, the thermalization condition 
requires a longer time. The other aspect of the figures 
worthy of mention is the difference between the behavior of 
$T_e^*$ and $t^*$ in the region of low values of $t_p$, for 
both $i_p = 4$ and $i_p=10$. While $t^*$ increases with $t_p$ 
up to reach a maximum value and then decreases, $T_e^*$ 
decreases monotonously. These last asymptotic behavior would 
correspond to the values of time and temperature of 
thermalization that the system would reach 
in the event of a CW laser mode experiment in 
which the same rate of generation is kept during a long time, 
i.e., $t_p \rightarrow \infty$. 
Again here the larger energy in excess received by the carriers
and the large energy of states of the system with $i_p=10$
explain the corresponding larger value of $T_e^*$ and $t^*$, in
comparison to the respective results for $i_p=4$.

The carrier population at the lowest level in the conduction 
band governed by Eq.\ (14.c) has an important role in the 
thermalization process\cite{marphasetr}. 
The shoulder shown in the behavior of 
$t^*$ as a function of the pulse duration $t_p$, is caused
predominantly by two aspects of the carrier kinetics. One of them
is the time necesary to form in the
lowest level a carrier population, which in the thermalized condition
must exponentially decay for increasing energies. The other 
aspect is the rate of carrier generation at the level of energy
$i_p$, which in the thermalized condition should be small enough in
order that this generation at each step becomes included in the error
tolerance in the fitting.

\begin{figure}
\includegraphics[width=2.75in]{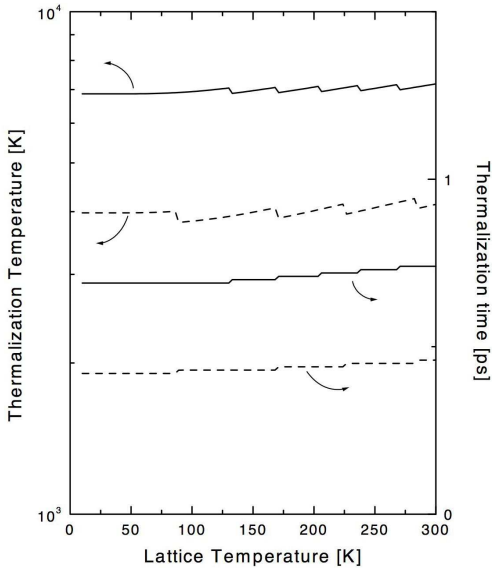}
\caption{$T_e^*$ and $t^*$ as a function of the lattice 
temperature $T_L$ for $N=10^{17}$ cm$^{-3}$ and two pulse 
durations $t_p=1$ ps (continuous lines) and $t_p=10$ ps 
(dashed lines).}
\end{figure}
Fig.\ 3 shows how the values of $t^*$ and $T_e^*$ change
with the lattice temperature $T_L$. We consider here two 
different pulse durations $t_p=1$ ps (continuous line) and 
$t_p=10$ ps (dashed line), an excitation energy $i_p=4$ 
and a carrier concentration $N=10^{17}$ cm$^{-3}$.
We observe that the lattice temperature has a rather mild 
effect on the thermalization process. According to the 
behavior of the ECFs during the thermalization process 
(Fig.\ 1), the major influence must come from phonon 
absorption events, whose contribution is larger for higher 
$T_L$ values, this is so, because it propitiates a high 
LO-phonon population in such a way that, comparatively to 
the phonon emission, phonon absorption increases its 
participation in the kinetics of the carrier system.   

\begin{figure}
\includegraphics[width=2.75in]{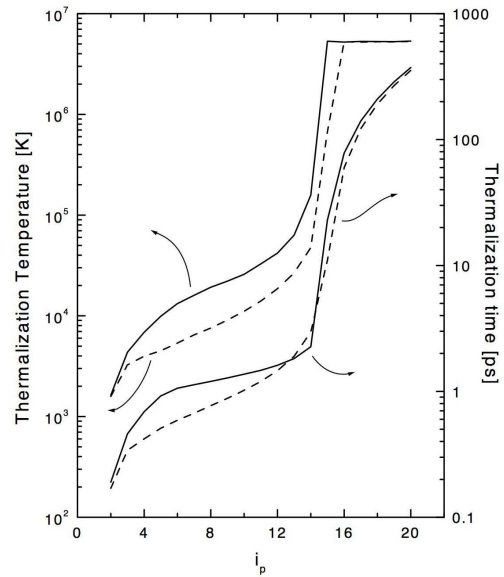}
\caption{$T_e^*$ and $t^*$ vs energy level of excitation $i_p$ 
for the same pulses as Fig.\ 3.}
\end{figure}
In Fig.\ 4, for the same pulses as in Fig.\ 3, we have 
depicted our results for $t^*$ and $T_e^*$ as a function of 
the energy of excitation $i_p$.  The dashed curves correspond 
to $t_p=10$ ps. The non-linear nature of the $e$-$e$ 
interaction is reveled in these results. We observe that a 
general behavior of both quantities is that, they increase 
with increasing values of the excitation energy, however, 
this behavior is non-monotonic. The dependence of the carrier 
temperature upon the carrier concentration and upon the energy 
of excitation have been studied since the 70's decade. We have
studied this dependences theoretically. In particular we 
have  found that strong changes in the CDF, associated with 
the non-linear nature of the $e$-$e$ interaction, can be 
induced by varying the excitation energy\cite{carrilloreyes}. 
The abrupt change in $T_e^*$ and $t^*$ about $i_p=14$ is a 
manifestation of this sensitive dependence. There has been in 
the literature some discussion regarding to this point. 
In particular, this 
phenomenon has been analized as a phase transition like 
behavior of the carrier population in the lowest level of 
energy\cite{marphasetr}. 

\begin{figure}
\includegraphics[width=2.75in]{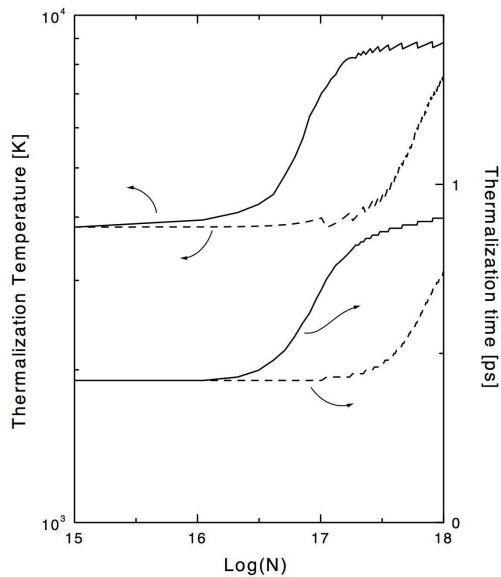}
\caption{Dependence of $T_e^*$ and $t^*$ on the carrier 
concentration. Energy of excitation $i_p=4$ and pulses of 
$t_p=1$ ps (continuous lines) and $t_p=10$ ps 
(dashed lines).}
\end{figure}
In Fig.\ 5 we show the dependence of $T_e^*$ and $t^*$ on the 
carrier concentration. We also consider here two pulses of 
duration $t_p=1$ ps, continuous lines, and $t_p=10$ ps, 
dashed lines, an energy level of excitation $i_p=4$, and a 
lattice temperature $T_L=10$ K. The non-linear dependence of 
$t^*$ and $T_e^*$ upon the carrier concentration is clearly 
exhibited. Notice that for carrier concentrations larger than 
$10^{17}$ cm$^{-3}$ the screening effects begin to be 
noticeable. Screening turns the $e$-$e$ interaction 
less effective in 
thermalizing the CDF and reduces the rate at which $e$-ph 
scattering takes out energy from the electronic system, in 
this way, both $T_e^*$ and $t^*$ increases with the carrier 
concentration. 

\section{Comments and remarks}
The criterion we have applied here to define $T_e^*$ and $t^*$ 
by means of the expression (\ref{p2000_eq_15}) may require some 
improvement. We think that the least squares fitting to an 
exponential function, although it resembles an experimental 
procedure, the included inherent error could be the 
origin
of some of the uneven behavior 
we observe in various of our results. However, the agreement 
of our results in the description of the cooling and the steady state 
processes in hot electron systems\cite{marjl2},
with the experimental data, 
provides 
support to our theoretical model to describe
the ultra-fast phenomena which lead a 
photo-generated carrier population to reach the thermalized 
condition. The reported experimental and theoretical 
results on the subject of the kinetics of the 
thermalization\cite{hess} also support our definition.
One of the advantages of our ``kinetic'' approach 
is that, due to the
fact that
ECFs depend only on the band structure of 
the system and consequently on the system dimensionality, it 
can be applied to study some ultra-fast phenomena in systems 
of reduced dimensionality and also systems with a small number 
of particles. In fact we have applied these theoretical 
framework to study some transport phenomena in mesoscopic 
semiconductor heterostructures\cite{lilia96}.  An additional 
interesting characteristic of this theoretical procedure is 
that, the required numerical calculations are not at all 
expensive. All the results we present in this paper can be 
obtained in a few minutes in an ordinary PC. In our analysis 
we have focused our attention to the role that the main interaction 
mechanisms play, however, the inclusion of some other 
scattering processes in the kinetics of the system is a simple 
task in this theoretical scheme\cite{marjl2}. 

In conclusion, we have
presented a detailed analysis of the thermalization process in terms of the
relevant external parameters, i.e., laser pulse duration, energy of the
photoexcitation, intensity of the photoexcitation, and
lattice temperature. We defined two physical parameters inherent to
the time evolution of the system on which
the analysis of thermalization have been done, the thermalization time and
temperature. We have found that the lattice temperature has a 
neglegible influence on $T_e^*$ and $t^*$. The other three external parameters
clearly influence the thermalization process. 

\acknowledgements
The partial financial support by CONACyT (M\'exico) is 
acknowledged.

\end{document}